# Electronic structure of Graphene/Co interfaces

D. Pacilé[1,2*], S. Lisi[3], I. Di Bernardo[3], M. Papagno[1,2], L. Ferrari[4], M. Pisarra[1], M. Caputo[1,5], S. K. Mahatha[6],
P. M. Sheverdyaeva[2], P. Moras[2], P. Lacovig[5], S. Lizzit[5], A. Baraldi[5,7,8], M. G. Betti[3], and C. Carbone [2]

[1] *Dipartimento di Fisica, Università della Calabria, 87036 Arcavacata di Rende (CS), Italy*
[2] *Istituto di Struttura della Materia, Consiglio Nazionale delle Ricerche, Trieste, Italy*
[3] *Dipartimento di Fisica, Università di Roma La Sapienza, Piazzale Aldo Moro 5, I-00185 Roma, Italy*
[4] *Istituto dei Sistemi Complessi, Consiglio Nazionale delle Ricerche, I-00133 Roma, Italy*
[5] *Elettra-Sincrotrone Trieste S.C.p.A., Strada Statale 14 Km 163.5, I-34149 Trieste, Italy*
[6] *International Center for Theoretical Physics (ICTP), I-34014 Trieste, Italy*
[7] *Dipartimento di Fisica, Università degli Studi di Trieste, Via Valerio 2, I-34127 Trieste, Italy*
[8] *IOM-CNR, Laboratorio TASC, Strada Statale 14 Km 163.5, I-34149 Trieste, Italy*

(Dated: September 14, 2014)

Photoemission, from core levels and valence band, and low-energy electron diffraction (LEED) have been employed to investigate the electronic and structural properties of novel graphene-ferromagnetic (G-FM) systems, obtained by intercalation of one mono-layer (1ML) and several layers (4ML) of Co on G grown on Ir(111). Upon intercalation of 1ML of Co, the Co lattice is resized to match the Ir-Ir lattice parameter, resulting in a mismatched G/Co/Ir(111) system. The intercalation of further Co layers leads to a relaxation of the Co lattice and a progressive formation of a commensurate G layer lying on top. We show the C *1s* line shape and the band structure of G in the two artificial phases, mismatched and commensurate G/Co, through a comparison with the electronic structure of G grown directly on a Co thick film. Our results show that while the G valence band mainly reflects the hybridization with the *d* states of Co, regardless of the structural phase, the C *1s* line shape is very sensitive to the rumpling of the G layer and the coordination of carbon atoms with the underlying Co. Even in the commensurate (1x1) G/Co phase, where graphene is in register with the Co film, from the angular dependence of the C *1s* core level we infer the presence of a double component, due to in-equivalent adsorption sites of carbon sub-lattices.

## I. INTRODUCTION

The synthesis of graphene (G) on transition metal surfaces allows to obtain high-quality large-area samples and to design and control the electronic properties through the interaction with other species. Adsorption of atoms, molecules and clusters on top, or intercalation beneath the graphene layer are two different strategies widely employed to this end [1]. The intercalation of noble metal atoms [2–4], or gaseous species [5–7] at the graphene-substrate interface, for instance, have been proven to soften the hybridization with the substrate and to lift the graphene layer. Alternatively, hydrogenation of G [8] or alkali metals deposition [9] have been used to open a wide electronic gap between $\pi$ and $\pi^*$ states of G, or to induce a superconductive behavior [10]. On the other hand, little has been done yet to manipulate the weak intrinsic magnetic properties of G, for instance by designing contacts with FM (Fe, Co, Ni), or by exploring the coupling with magnetic species [11–15]. Here, we focus our attention on the electronic structure of some selected G/FM systems. In order to grow G on FM, carbon segregation or chemical vapor deposition (CVD) of hydrocarbons on single crystals have been used so far [16–18]. Alternatively, single layers of Ni or Co atoms have been intercalated underneath G grown on Ir(111)[19, 20], with the benefit of tailoring the structural properties at the interface. The bare G/Ir(111) system has been studied with several methods, all describing a gently rippled moiré superstructure where carbon atoms interact very weakly with the underlaying metal, thus keeping an almost unperturbed graphene $\pi$ band [21]. The intercalation of a single epitaxial layer of Ni or Co [19, 20] leads to a locally enhanced interaction, resulting in a strong corrugation of the graphene layer, with a modulation in the range of 1.2-1.8 Å, and a minimum G-FM distance of about 2 Å. Based on scanning tunneling microscopy (STM) and density functional theory (DFT) results [19, 20], a pseudomorphic arrangement of intercalated Co (or Ni) atoms on Ir(111) with comparable interplane distance was deduced. The band structure of G/1MLNi/Ir(111) probed by angle-resolved photoemission spectroscopy (ARPES) shows a clear transition from a nearly free-standing to a strongly hybridized character of the $\pi$ band [19]. The hybridization between Co (or Ni) $d$ states and graphene $\pi$ states is directly related to the strongly interacting top-hollow and bridge configuration in the lower parts of the moiré mesh, while the interaction is van der Waals-like for other regions (fcc-hcp configurations). Interestingly, the moiré pattern is reflected into the magnetic properties also. Spin-polarized STM and DFT calculations revealed that the magnetization of G is site-dependent in G/1MLCo/Ir(111) [20]: in the hills (fcc-hcp configurations) G is ferromagnetically coupled to the Co underneath, whereas in the valleys (top-hollow) is coupled antiferromagnetically.

In order to explore the interaction mechanism of G-FM systems, we provide here a characterization of the electronic structure of G/Co. We design three different G/Co systems: the mismatched G/1MLCo/Ir, the commensurate (1x1), and the G/4MLCo/Ir intermediate phase. The comparison between photoemission spectra allow us to discuss the influence of lattice mismatch and chemical interaction, and thus to rank their effect on the graphene electronic structure.

---

*Email address: daniela.pacile@fis.unical.it

## II. METHODS

Angle-resolved photoemission and high resolution core level measurements were performed at VUV and SuperESCA beamlines of the Elettra synchrotron radiation facility (Trieste, Italy).

Experimentally, the G/Ir(111) system was prepared by exposing the Ir(111) surface at 1300 K to ethylene [22]. Intercalation of Co underneath a graphene layer was performed via annealing of the predeposited film in the temperature range of 600-800 K. Starting from the submonolayer regime, the Co coverage was estimated on bare Ir(111) by measuring the intensity ratio of Co $3p$ and Ir $4f$ core levels. The intercalation of 1ML of Co underneath G/Ir(111) was monitored following the evolution of Ir $4f$, Co $3p$ and C $1s$ core level photoemission spectra, acquired with photon energies of 176 eV and 400 eV. Cobalt adatoms deposited on the graphene layer do not influence the C $1s$ and Co $3p$ lineshape, suggesting the absence of Co-C intermixing (see Supplementary information). Moreover, in order to check possible Co-Ir mixing due to the annealing, the intensity ratio of Co $3p$ and Ir $4f$ was monitored. We do not detect (data not shown) the formation of Ir-Co surface alloy in the temperature range used here.

Thick Co film was grown on W(110) by evaporation of about 15 ML of Co on the clean substrate at room temperature. The graphene layer was grown by CVD of ethylene at about 700 K, at the lower limit of temperature to avoid local breaking of the Co film. Surface order and cleanliness was checked by LEED and core levels measurements. ARPES and core level measurements on Co thick films were taken at the VUV beamline.

At SuperESCA, photoelectrons were collected at normal emission, with a Phoibos electron energy analyzer equipped with a homemade delay line detection system. The incidence angle was 70°. Photoemission data have been normalized to the beam intensity, and the binding energy (BE) scale was calibrated using the Fermi edge of the Ir substrate.

ARPES data taken at VUV were collected using a Scienta R4000 electron energy analyzer, at the photon energy of 63 eV, where the contrast between C and Co related bands was optimal. Energy and angular resolution were 50 meV and 0.1°, respectively. All ARPES spectra here shown were acquired at room temperature.

## III. RESULTS AND DISCUSSION

### A. XPS

During the intercalation of Co underneath G/Ir the lineshape of the C $1s$ peak was recorded as a function of temperature, and displayed in Fig.1a. A two-dimensional (2D) projected plots of fast-XPS spectra is shown in the lower panel of Fig.1a. The C $1s$ profile upon completion of the intercalation process (Fig. 1b) has been fitted with two Doniach-Sunjic (DS) lineshapes [23]. The fit results, including BE, full width half maximum (FWHM), for both the lorentian and the gaussian contribution, and the asymmetry factor $\alpha$, are reported in the figure caption.

The C $1s$ peak of G/Ir(111) exhibits a sharp profile centered at 284.13 eV [24] (see Supplementary information). In Fig.1a, C $1s$ peak intensity stays constant up to about 570 K. Above this temperature, an almost complete drop of intensity was observed, while a double peaked line profile arises at higher binding energy, with components at 284.92 eV (A'), and 284.42 eV (B') (Fig.1b). We attribute the intensity drop to the intercalation of 1ML of Co underneath the G layer. For 1ML of Co, the C $1s$ G/Ir peak is completely quenched, and the A' peak area contributes by 74% to the total area, while the 26% is due to the B' peak. The LEED pattern reported in the insert of Fig.1b shows the moiré pattern with a reduced intensity of extra-spots with respect to G/Ir (not shown). Graphene grown on Ir(111) surface exhibits a characteristic moiré pattern with a periodicity of 2.5 nm due to the lattice mismatch and a slightly corrugation of the graphene overlayer with top-valley height of about 0.3 Å[25]. Recent studies confirm a moiré pattern with the same periodicity when a single Co layer is intercalated [20], but with a more pronounced corrugation, *i. e.* the height of carbon atoms in the top-region compared with the valley regions is 1.2-1.8 Å. Both corrugation and hybridization of the G layer are reflected in the C $1s$ core level line-shape. This structural configuration justifies the presence of a double peak binding energy distribution, as already observed for other corrugated G layers grown on selected metallic surfaces, like Ru(0001), Rh(111) and Re(0001) [26–28]. The degree of corrugation is not only related to the lattice mismatched substrates, but critically depends also on the strength of interaction of the G layer with the underlying metal surface. The double component can be associated to a corrugated G layer as sketched in Fig. 1b, where the peak with lower binding energy B' is due to weakly interacting carbon atoms, while the high-energy peak A' results from highly interacting G atoms.

In the parental G/1MLNi/Ir(111) system a main component centered at 284.90 (±0.20) eV was measured [19], with a total width of about 840 meV and a tail toward lower binding energy. Considering all carbon adsorption sites, the theoretical simulation has shown that the strongly interacting C atoms in the lower regions are dominant, while all the other contributions are spread at lower binding energy with respect to the main peak. The minimum G-Co(Ni) distance was predicted at about 2 Å, which indeed gives a consistent binding energy of the main (A') peak.

A subsequent deposition of 3ML of Co on top of G can induce further intercalated layers, following the same annealing procedure. The C $1s$ core level lineshape evolves as a function of temperature, and the binding energy exhibits a slight progressive shift, as reported in Fig. 2a. At the completion of 4 ML of Co, the C $1s$ peak exhibits a wide and asymmetric line-shape, suggesting a different structural configuration of the G layer on the Co film. A qualitative confirm of a differ-ent arrangement can be deduced from the diffraction pattern of G/4ML Co/Ir, reported in the inset of Fig. 2b. Here, the LEED pattern reveals an almost complete smearing out of the moiré superstructure, suggesting a release of the mismatch. During the intercalation of several layers, the Co film progressively releases its in-plane lattice constant and corrugation at the interface, until the G layer becomes in register. The C 1s lineshape reported in Fig. 2b is due to several non-equivalent C configurations on a Co substrate not fully relaxed, as deduced further on from the comparison with the Co film.

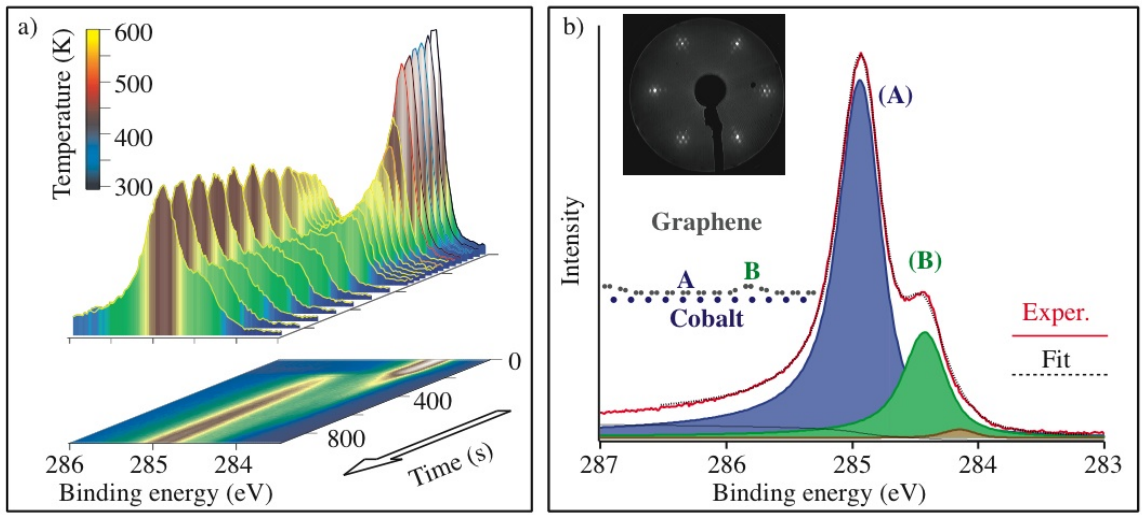

FIG. 1: a) Fast XPS spectra of the C1s core levels during the intercalation of 1 ML of Co under G/Ir(111) as a function of temperature, in 1000s, shown as two-dimensional projected plot (lower panel) and intensity line profiles (upper panel). The color of each profile is related to the sample temperature, as reported in the color scale. b) C1s core levels for G/1ML Co/Ir(111) fitted as the sum of DS line profiles. The two main contributions are due to carbon atoms in valleys (A′ at 284.92 eV, with a gaussian FWHM of 0.21 eV, a lorentian FWHM of 0.25 eV and an $\alpha$ asymmetry of 0.10) and hills (B′ at 284.42 eV, with a gaussian FWHM of 0.21 eV, a lorentian FWHM of 0.25 eV and an $\alpha$ asymmetry of 0.10), as sketched in the lower inset. In the upper inset, the LEED (140eV primary energy) is reported: a moiré superstructure, arising from the Co-C lattice mismatch is visible.

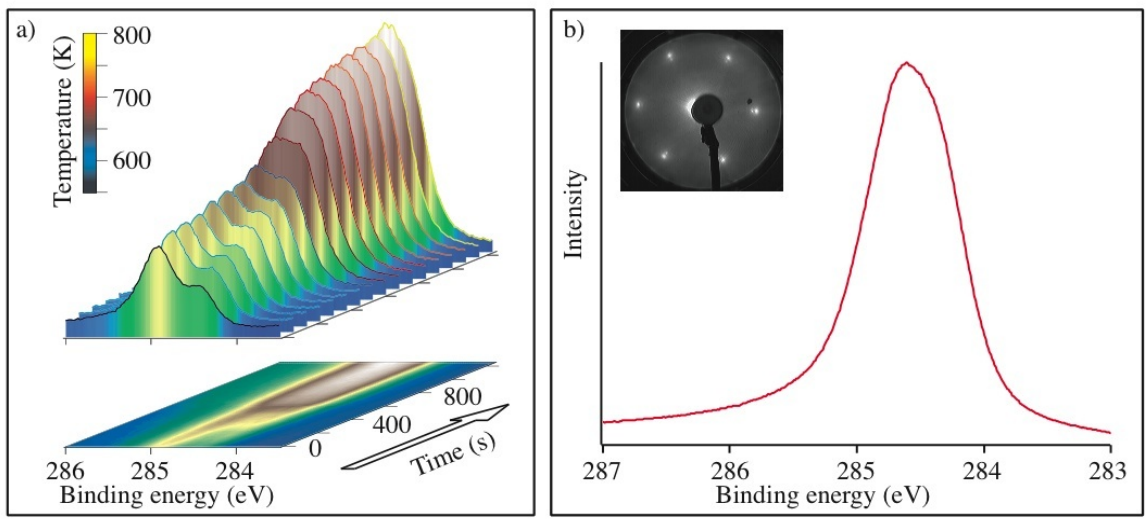

FIG. 2: a) Fast XPS spectra of the C1s core levels during the intercalation of 3 Co MLs under G/1ML Co/Ir(111) as a function of temperature, in 1200s, shown as two-dimensional projected plot (lower panel) and intensity line profiles (upper panel). The color of each profile is related to the sample temperature, as reported ind the color scale. b) C1s core levels for G/4ML Co/Ir(111). In the upper inset, the LEED (140eV primary energy) is reported.

To shed more light on the overall interaction of graphene with several Co layers, we have performed photoemission measurements on G grown on thick films (15ML) of Co(0001) on W(110). Contrary to previous measurements [18], we were able to achieve a single domain growth of G on Co, as shown by LEED (inset of Fig. 3a) and by ARPES results discussed afterwards. Figure 3a shows the angularly integrated C 1$s$ core level of G grown on the Co film. Even when G is in register with the Co substrate, the two carbon atoms occupy non-equivalent adsorption sites. The crystal structures of both G/Ni and G/Co are well established by quantitative (*I-V*) low-energy electron diffraction [30], STM [18, 31, 32], and DFT [31, 33], all suggesting a top-hollow configuration for the two carbon sub-lattices (inset of Fig. 3a). Recently, a top-bridge configuration of G on Ni(111) was proposed [34], together with stochastic structural phase transitions during the cooling of the sample. In G/Co/W, no change of the line-shape was seen as a function of temperature. Considering the strong chemical interaction between G and Ni (or Co), and a small G-FM separation (of about 2 Å), a sub-lattice asymmetry is expected for both top-bridge and top-hollow configurations. The C 1$s$ peak of Fig. 3a exhibits an experimental FWHM of about 700 meV. The photoelectron diffraction map of the C 1$s$ peak (taken with angular resolution of 0.3°) reveals a periodic modulation of the FWHM, reported in Fig. 3b, with an excursion between maximum (bright yellow) and minimum (black) values of about 120 meV. While traveling through the crystal, the photoelectron is backscattered at the periodic crystal potential leading to wavelength modulations of the final-state electron wave function. The line-width modulation here seen must be related to diffraction effects arising from different sites of the C 1$s$ photoemission process, ruling out the existence of one single component. Thus, we conclude that the C1$s$ line-shape reported in Fig. 3a is representative of a stable (1x1) structural phase, where the two carbon sub-lattices occupy non-equivalent adsorption sites, with a separation between their components within 100 meV.

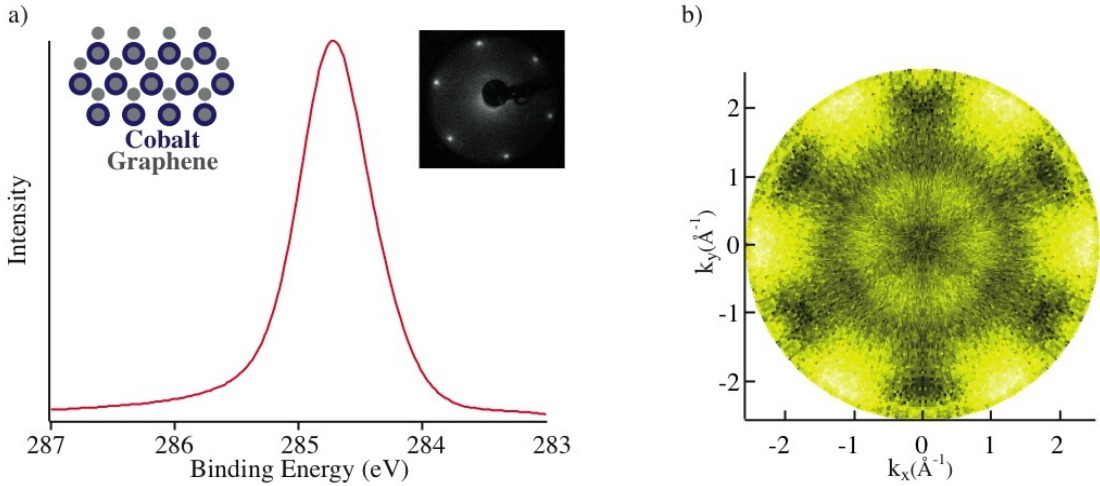

FIG. 3: G/Co/W(110): (a) C 1$s$ core level taken with photon energy of 400 eV. In the inset: (left) example of a structural model (top-hollow) where carbon atoms occupy inequivalent adsorption sites; (right) LEED (122 eV primary energy) showing a single domain (1x1) G/Co. (b) $k$-space image of the C 1$s$ FWHM. Bright yellow corresponds to maximum values of FWHM.

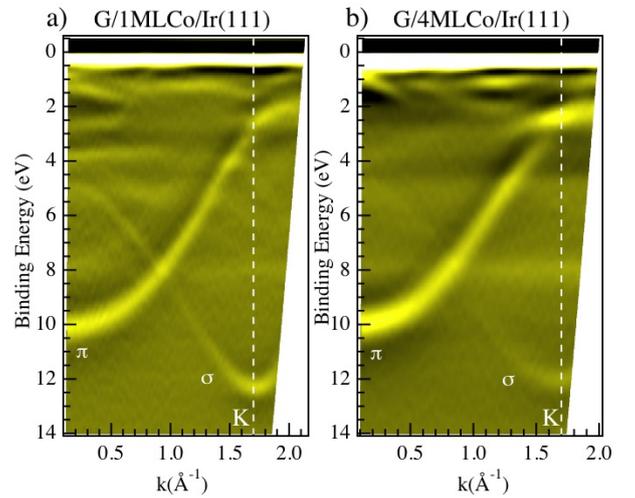

FIG. 4: (Color online) ARPES dispersions shown in second derivative along ΓK as a function of Co atoms intercalated underneath G/Ir(111): (a) 1ML Co; (b) 4ML Co.

## B. ARPES

In order to investigate the effect of lattice mismatch on the valence bands, ARPES has been employed to study G/Co in different structural phases. Fig. 4 shows ARPES maps of the electronic band dispersion of (a) G/1MLCo/Ir(111) and (b) G/4MLCo/Ir(111), taken along the ΓK direction. In Fig. 4a the $\pi$ state of graphene exhibits a minimum at the Γ point at 10.10 eV, and it merges with the $d$ states of Co at 2.50 eV at the K point, where the planar $\sigma$ state reaches 12.20 eV of binding energy. When several monolayers of Co are intercalated via annealing (Fig. 4b) $d$ bands of Co are enhanced while those of Ir are barely detected. Nevertheless, one clearly sees that the $\pi$ band is almost unaffected along the ΓK direction. Similar results have been found for the parental G/Ni/Ir(111) system [19], where the $\pi$ band behavior is marginally affected by the number of intercalated Ni layers.

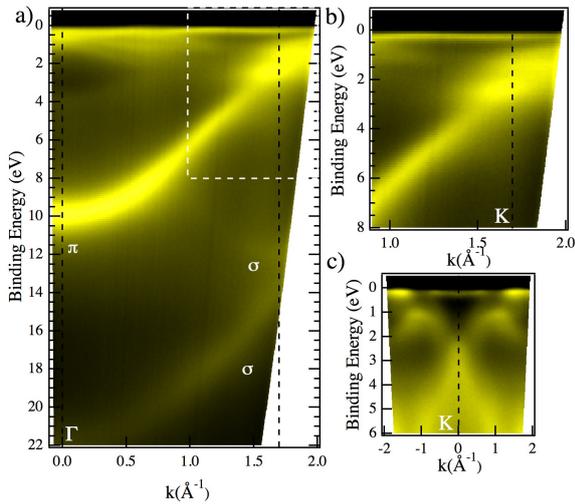

FIG. 5: (Color online) Photoemission measurements of G/Co/W(110). (a) ARPES dispersion sampled along ΓK, taken with photon energy of 63 eV; (b) magnified region of interest enclosed in the white dashed area in (a); (c) extracted ARPES dispersion along the direction perpendicular to ΓK.

Fig. 5a shows the overall band structure of G/Co/W(110) sampled along ΓK. According to the literature, and similarly to the intercalated systems here shown, the $\pi$ band appears shifted to higher binding energy by the interaction with the substrate with respect to freestanding graphene, with a minimum at about 10.10 eV at Γ. To see the two branches of the $\pi$ band, thus avoiding the destructive interference of photo-electrons, we have sampled the dispersion along the direction perpendicular to ΓK (Fig. 5c). Our data suggest that the hybridization takes place between the $d$ states of Co and both $\pi$ and $\pi^*$ bands, with a crossing point at about 2.80 eV, in agreement with previous measurements [18]. The position of the maximum of the $\pi$ band is downshifted by 0.30 eV with respect to the intercalated systems. This difference could be related to the width of the Co-$d$ states as a function of the Co thickness, and thus to the degree of hybridization with the $\pi$ bands, in analogy with G/Ni/Ir(111) [19]. Finally, in Fig. 5b we have magnified the region around the K point of Fig. 5a (enclosed by the white dashed line). One clearly sees that no additional contributions which would come from different domains are present. The growth of graphene on bulk Co or Ni is then similar, as can be deduced by considering the close lattice mismatch of 1.2% (G-Ni) and 1.9% (G-Co).

## IV. CONCLUSION

We have studied the G-Co interaction in two different structural phases, mismatched and commensurate (1x1), while keeping the chemical environment similar. We show that the C 1$s$ core level is very sensitive to the coordination of carbon atoms, namely to the position of the two graphene sub-lattices, and further to the corrugation of the graphene layer. In the mismatched G/Co/Ir phase, with Co bonds stretched to the Ir lattice constant, the C1$s$ peak exhibits two main components related to different carbon adsorption sites on valleys and hills, in analogy with strongly corrugated systems. In the commensurate (1x1) phase the C 1$s$ line-width is of about 700 meV, with a modulation of 120 meV demonstrating the existence of more than one single component. We conclude that going from the mismatched to the commensurate phase, the sub-lattice asymmetry is always reflected into the C 1$s$ core level. The binding energy spread of C 1$s$ is much more pronounced when the graphene layer is corrugated, thus the two sub-lattices occupy a high number of non-equivalent adsorption sites. On the other hand, the G-Co chemical interaction is dominant in the valence band, regardless of the structural phase. In G/1MLCo/Ir no $\pi$ band splitting is observed for valleys and hills, at odd with $h$-BN monolayers grown on selected transition metals [36]. This is due to the metallic nature of graphene [35], which prevents charge confinement effects observed in $h$-BN.

[1] M. Batzill, Surf. Sci. Rep. **67**, 83 (2012).
[2] A. Varykhalov, J. Sánchez-Barriga, A. M. Shikin, C. Biswas, E. Vescovo, A. Rybkin, D. Marchenko, and O. Rader, Phys. Rev. Lett. **101**, 157601 (2008).
[3] A. Varykhalov, M. R. Scholz, T. K. Kim, and O. Rader, Phys. Rev. B **82**, 121101 (2010).
[4] M. Papagno, P. Moras, P. M. Sheverdyaeva, J. Doppler, A. Garhofer, F. Mittendorfer, J. Redinger, C. Carbone, Phys. Rev. B **88**, 235430 (2014).
[5] H. Zhang, Q. Fu, Y. Cui, D. Tan, and X. Bao, J. Phys. Chem. C **113**, 8296 (2009).
[6] R. Larciprete, S. Ulstrup, P. Lacovig, M. Dalmiglio, M. Bianchi, F. Mazzola, L. Hornekaer. F. Orlando, A. Baraldi, P. Hofmann, and S. Lizzit, ACS Nano **6**, 9551 (2012).
[7] S. Lizzit, R. Larciprete, P. Lacovig, M. Dalmiglio, F. Orlando, A. Baraldi, L. Gammelgaard, L. Barreto, M. Bianchi, E. Perkins, and P. Hofmann, Nano Lett **12**, 4503 (2012).
[8] D. Haberer et al., Nano Lett. **10**, 3360 (2010).
[9] M. Papagno, S. Rusponi, P. M. Sheverdyaeva, S. Vlaic, M. Etzkorn, D. Pacilé, P. Moras, C. Carbone, and H. Brune, ACS Nano **6**, 199 (2012).
[10] Z.-H. Pan, J. Camacho, M. H. Upton, A. V. Fedorov, C. A. Howard, M. Ellerby, and T. Valla, Phys. Rev. Lett. **106**, 187002 (2011).
[11] M. Weser, E. N. Voloshina, K. Horn, and Y. S. Dedkov, Phys. Chem. Chem. Phys. **13**, 7534 (2011).
[12] S. Bhandary, O. Eriksson and B. Sanyal, Sci. Reports **3**, 3405 (2013).
[13] M. Bazarnik, J. Brede, R. Decker, and R. Wiesendanger, ACS Nano **7**, 11341 (2013).

[14] S. Vlaic, A. Kimouche, J. Coraux, B. Santos, A. Locatelli, and N. Rougemaille, Appl. Phys. Lett. **104**, 101602 (2014).
[15] O. V. Yazyev, A. Pasquarello, Phys. Rev. B **80**, 035408 (2009).
[16] C. Oshima and A. Nagashima, J. Phys. Condens. Matter **9**, 1 (1997).
[17] E. Voloshina and Yu. Dedkov, Phys. Chem. Chem. Phys. **14**, 13502 (2012).
[18] A. Varykhalov, D. Marchenko, J. Sánchez-Barriga, M. R. Scholz, B. Verberck, B. Trauzettel, T. O. Wehling, C. Carbone, and O. Rader, Phys. Rev. X **2**, 041017 (2012).
[19] D. Pacilé, P. Leicht, M. Papagno, P. M. Sheverdyaeva, P. Moras, C. Carbone, K. Krausert, L. Zielke, M. Fonin, Y. S. Dedkov, F. Mittendorfer, J. Doppler, A. Garhofer, and J. Redinger, Phys. Rev. B **87**, 035420 (2013).
[20] R. Decker, J. Brede, Nicolae Atodiresei, V. Caciuc, S. Blugel, and R. Wiesendanger, Phys. Rev. B **87**, 041403(R) 2013.
[21] I. Pletikosić, M. Kralj, P. Pervan, R. Brako, J. Coraux, A. T. N'Diaye, C. Busse and T. Michely, Phys. Rev. Lett. **102**, 056808 (2009).
[22] S. Rusponi, M. Papagno, P. Moras, S. Vlaic, M. Etzkorn, P. M. Shverdyaeva, D. Pacilé, H. Brune, and C. Carbone, Phys. Rev. Lett. **105**, 246803 (2010).
[23] S. Doniach, and M. Sunjic, J. Phys. C **3**, 285 (1970).
[24] S. Lizzit, G. Zampieri, L. Petaccia, R. Larciprete, P. Lacovig, E. D. L. Rienks, G. Bihlmayer, A. Baraldi and P. Hofmann, Nat. Phys. B **6**, 345 (2010).
[25] A. T. N'Diaye, J. Coraux, T. N. Plasa, C. Busse, and T. Michely, New J. Phys. **10**, 043033 (2008).
[26] A. B. Preobrajenski, M. L. Ng, A. S. Vinogradov, and N. Mårtensson, Phys. Rev. B **78**, 073401 (2008).
[27] E. Miniussi, M. Pozzo, A. Baraldi, E. Vesselli, R. R. Zhan, G. Comelli, T. O. Mentes, M. A. Niño, A. Locatelli, S. Lizzit, and D. Alfè, Phys. Rev. Lett. **106**, 216101 (2011).
[28] D. Alfè, M. Pozzo, E. Miniussi, S. Günther, P. Lacovig, S. Lizzit, R. Larciprete, B. Santos Burgos, T. O. Menteş, A. Locatelli and A. Baraldi, Sci. Rep **3**, 2430 (2013).
[29] C. Busse, P. Lazic, R. Djemour, J. Coraux, T. Gerber, N. Atodiresei, V. Caciuc, R. Brako, A. T. N. Diaye, S. Blügel, J. Zegenhagen, and T. Michely, Phys. Rev. Lett. **107**, 036101 (2011).
[30] Y. Gamo, A. Nagashima, M. Wakabayashi, M. Terai, and C. Oshima, Sur. Sci. **374**, 61 (1997).
[31] D. Eom, D. Prezzi, K. T. Rim, H. Zhou, M. Lefenfeld, S. Xiao, C. Nuckolls, M. S. Hybertsen, T. F. Heinz, and G. W. Flynn, Nano. Lett. **9**, 2844 (2009).
[32] L. V. Dzemiantsova, M. Karolak, F. Lofink, A. Kubetzka, B. Sachs, K. von Bergmann, S. Hankemeir, T. O. Wehling, R. Fromter, H. P. Oepen, A. I. Lichtenstein, and R. Wiesendanger, Phys. Rev. B **84**, 205431 (2011).
[33] P. A. Khomyakov, G. Giovannetti, P. C. Rusu, G. Brocks, J. van den Brink, and P. J. Kelly, Phys. Rev. B **79**, 195425 (2009).
[34] W. Zhao, S. M. Kozlov, O. Höfert, K. Gotterbarm, M. P. A. Lorenz, F. Vines, C. Papp, A. Görling, and Hans-Peter Steinrück, J. Phys. Chem. Lett. **2**, 759 (2011).
[35] T. Brugger, S. Günther, B. Wang, J. H. Dil, M.-L. Bocquet, J. Osterwalder, J. Wintterlin, and T. Greber, Phys. Rev. B **79**, 045407(2009).
[36] T. Greber, M. Corso, and J. Osterwalder, Surf. Sci. **603**, 1373 (2009).

**Acknowledgements**

This work has been supported by MIUR (FIRB-Futuro in Ricerca 2010- Project PLASMOGRAPH Grant No. RBFR10M5BT, and PRIN grant 20105ZZTSE), and by the European Science Foundation (ESF) under the EUROCORES Program EuroGRAPHENE.